# TESTS OF THE EXTENDED RANGE SRF CAVITY TUNERS FOR THE LCLS-II-HE PROJECT*

C. Contreras-Martinez†, T. Arkan, A. Cravatta, B. Hartsell, J. Kaluzny, T. Khabiboulline, Y. Pischalnikov, S. Posen, G. Romanov, JC. Yun, Fermilab, Batavia, IL, USA

*Abstract*

The LCLS-II HE superconducting linac can produce multi-energy beams by supporting multiple undulator lines simultaneously. This could be achieved by using the cavity SRF tuner in the off-frequency detune mode. This off-frequency operation method was tested in the verification cryomodule (vCM) and CM 1 at Fermilab at 2 K. In both cases, the tuners achieved a frequency shift of -565±80 kHz. This study will discuss cavity frequency during each step as it is being assembled in the cryomodule string and finally when it is being tested at 2 K. Tracking the cavity frequency helped enable the tuners to reach this large frequency shift. The specific procedures of tuner setting during assembly will be presented.

## INTRODUCTION

The LCLS-II-HE project will add 19 cryomodules of the type already used for the LCLS-II project. For multi-energy operation in the LCLS-II-HE linac the tuners were modified to meet the off-frequency operation (OFO) specification. This new mode of operation requires the tuner to be able to compress the cavity by -465 kHz from the nominal operation of 1.3 GHz. The OFO is then 1299.535 MHz.

The SRF tuner for the LCSL-II-HE must be capable of bringing 100 % of all cavities to the operational frequency of 1.3 GHz. In the case of OFO at least 62 % of the cavities must be tuned to 1299.535 MHz [1]. Tuning from 1.3 GHz to OFO must be done approximately twice a month. This level of operation pushes the tuner and cavity to new thresholds, which test the longevity of both. Recent results demonstrate that the cavity and tuner can achieve these requirements [2]. The tuner needed to be modified to achieve these results. This paper discusses the successful tuning to OFO (1.3 GHz – 465 kHz) for the vCM and CM1 LCLS-II-HE cryomodules.

## CAVITY FREQUENCY TUNER

The tuner for LCLS-II-HE cavities consists of two components, one is the slow-coarse component and the other is the fine-fast component. The slow-coarse component consists of a Phytron stepper [3]. This component is used to tune the cavity to the nominal frequency after cooldown. The second frequency tuning component consists of two piezo actuator encapsulations made by Physik Instrumente (PI) used for fast-fine frequency [3].

In addition to the -465 kHz compression from 1.3 GHz the tuner must also compress the cavity from the 2 K landing frequency ($f_{2k\ Landing}$). The $f_{2k\ Landing}$ is the initial frequency after the cavities reach 2 K and no tuning has been performed. The next section will discuss how cavities arrive at this frequency. The results from [4] showed that 95 % of the cavities $f_{2k\ Landing}$ are below 250 kHz. Therefore, the tuner must compress the cavity by -715 kHz, roughly three times larger than the LCLS-II tuners. The LCLS-II-HE tuners were modified from the LCLS-II by decreasing the tuner lever arm ratio from 1:20 to 1:16. The length of the motor arm was also increased by 7 mm. Not all the changes were implemented for cavity one. A full description and discussion of these changes is presented in [4]. These modifications changed the motor sensitivity from 1.4 Hz/step to 1.84 Hz/step.

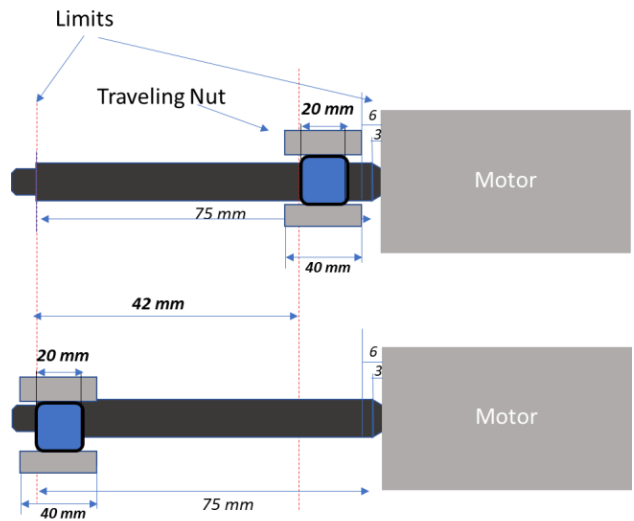

Figure 1: Motor shaft schematic showing the traveling range in mm. the traveling nut is connected to the tuner arm which is responsible for compressing the cavity.

The stepper motor system consists of two limit switches. They stop the motor when they are triggered. Three different parts can set the motor's hard limits. One of the limits is the length of the shaft screw thread, which is 75 mm (see Fig. 1). This limit is critical, if the traveling nut comes out of the threaded screw no movement will be possible. The second limit is the magnetic shielding on the cavity to the right of the end of the shaft screw. Making contact with the magnetic shielding will cause large forces on the stepper motor if it is operated. Large forces can lead to stepper motor failure. The last limit is when the traveling nut frame is too close to the motor body. The limit towards the body of the motor has two different gaps set to prevent hitting the motor body, as shown in Fig 1. The larger gap of 6 mm is only for cavity 1, and for the rest of the cavities it is 3 mm. Cavity 1 is different since not all changes made to the tuner were applied to to this cavity. This tuner has the additional function of supporting the gate vale, making modification difficult. A 3 mm gap is also set between the limit switch and the hard limit. This gap is set in case the cavity frequency is below 1.3 GHz when it is at 2 K. In the case of cavity 1 the full range of movement is 39 mm (assuming no other interference); for the rest of the cavities, it is 42 mm. The titanium shaft screw shrinks 0.11 mm from room temperature to 2 K (considering the 75 mm threaded length). This simple and simplified shrinkage of the shaft screw for cavity 1 results in an estimated tuning range of -717.6 kHz. The estimated tuning range for the rest of the cavities is -772.8 kHz. Note that this range is only attainable if no other interference is discovered.

## CAVITY FREQUENCY HISTORY

Another way to reach the full range of the OFO operation is to decrease the $f_{2k\ Landing}$. This depends on how the cavity is handled during processing at Fermilab and the manufacturer site. As discussed earlier, 95 % of the cavities tested had a $f_{2k\ Landing}$ of 250 kHz or less, if the $f_{2k\ Landing}$ is smaller the tuner will be able to reach full OFO range for more cavities. So far, two cryomodules have been tested at Fermilab, the vCM and CM 1. The cavity π frequency history for these two cryomodules is shown in Figures 2 and 3. In these figures, the frequency is measured at the manufacturer site. The unrestrained cavity frequency (cavity is without safety brackets) and the $f_{2k\ Landing}$ frequencies are measured at FNAL.

The frequency measurements shown in Figures 2 and 3 are done at different pressures and temperatures. There are three regions where the pressure will affect the cavity frequency. These are the pressure outside the helium vessel ($P_{out}$), the pressure in the cavity helium vessel ($P_{HeVessel}$), and in the cavity beamline ($P_{BeamLine}$). The different states where the cavity frequency is measured are shown in Table 1. Case 1 is the measurement at the manufacturer site. Case 2 is the unrestrained frequency measurement at FNAL. Case 3 is the measurement done at 2 K in FNAL.

Table 1: Cavity frequency, temperature, and pressure at different stages.

|        | T [K] | $P_{out}$ | $P_{HeVessel}$ | $P_{BeamLine}$ |
|--------|-------|-----------|----------------|----------------|
| Case 1 | 293   | 1 atm     | 1 atm          | Vacuum         |
| Case 2 | 293   | 1 atm     | 1 atm          | Vacuum         |
| Case 3 | 2     | Vacuum    | 23 Torr        | Vacuum         |

Note that the pressures at 2 K differ from those at room temperature. Using the cavity frequency difference between the 2 K landing and the unrestrained values ($f_{diff}$) the 2 K landing for future cavities can then be estimated from the unrestrained frequency. The average frequency difference $f_{diff}$ from the two cryomodules is 1.8453 MHz and the standard deviation is 0.0245 MHz. The $f_{2k\ Landing}$ can then be calculated with 95 % confidence (treating the values as a normal distribution) by $f_{unrestrained}$ + (1.8453+2*0.0245) MHz.

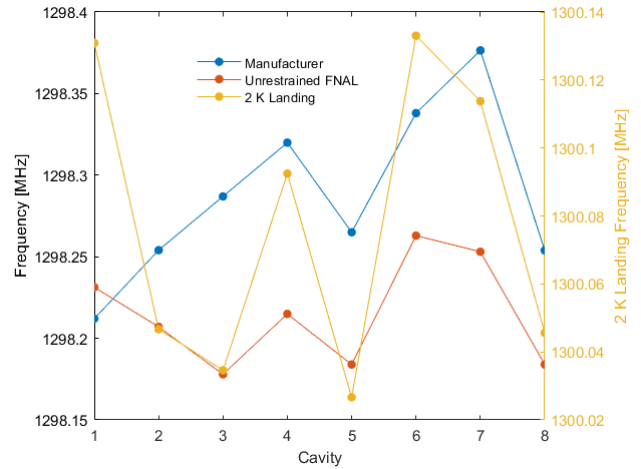

Figure 3: Cavity history of the CM 1 cavities.

Table 2: Steps required to tune the cavity to 1.3 GHz

| Cav. | vCM Steps to 1.3 GHz | CM1 Steps to 1.3 GHz |
|------|----------------------|----------------------|
| 1    | 40125                | 73500                |
| 2    | 18875                | 26360                |
| 3    | 7768                 | 19700                |
| 4    | 14885                | 52325                |
| 5    | 34145                | 15212                |
| 6    | 52330                | 77320                |
| 7    | 30300                | 63245                |
| 8    | 39500                | 25727                |

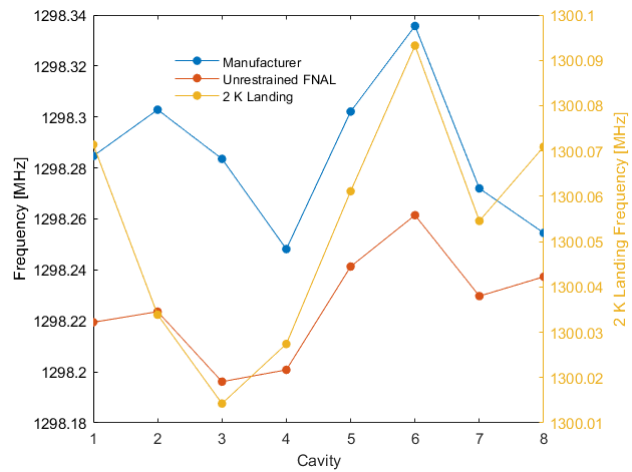

Figure 2: Cavity history of the vCM cryomodule. The frequency measured by the manufacturer, unrestrained at FNAL, and the 2 K landing frequency is shown.

Table 2 shows the number of steps required to tune the cavity to 1.3 GHz from the 2 K landing frequency. The further away the 2 K landing frequency is from 1.3 GHz the more steps are needed to tune the cavity. The motor sensitivity of roughly 1.84 Hz/ step is based on the total number of steps traveled. Figures 2 and 3 show that none of the cavities were above 1300.250 MHz, which means that all cavities should reach the OFO range. In the next section, it is shown that this is achieved.

## TUNERS TO OFO

The total number of steps from the 2 K landing frequency to the OFO set point is given in Table 3. For the vCM, only a fixed number of steps were done from 1.3 GHz. No limit switches were tripped during this operation. For CM1, the number of steps needed to reach OFO was larger due to the larger 2 K landing frequencies. During OFO operation none of the limit switches were tripped for CM1. It took a total of 1.5 hours to bring all cavities from 1.3 GHz to the OFO setpoint and then back to 1.3 GHz. The traveling nut travel length is given in Table 3. Note that the distance is much smaller than the estimated values calculated above.

Table 3: Steps required to tune the cavity to OFO frequency. This is the total number of steps since the 2 K landing frequency.

| Cav. | vCM OFO Steps | $vCM$ Shaft L [$mm$] | CM1 OFO Steps | $CM1$ Shaft L [$mm$] |
|---|---|---|---|---|
| 1 | 298125 | 29.81 | 335977 | 33.60 |
| 2 | 276875 | 27.69 | 290695 | 29.07 |
| 3 | 265768 | 26.58 | 284445 | 28.44 |
| 4 | 272885 | 27.29 | 316605 | 31.66 |
| 5 | 292145 | 29.21 | 280037 | 28.00 |
| 6 | 310330 | 31.03 | 356820 | 35.68 |
| 7 | 288300 | 28.83 | 325595 | 32.56 |
| 8 | 297500 | 29.75 | 287682 | 28.77 |

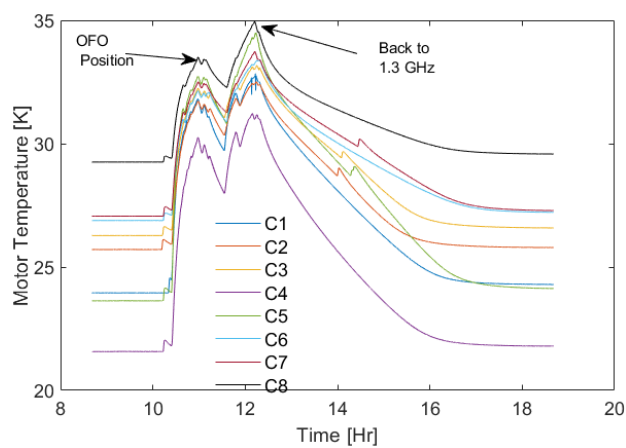

Figure 4: Motor temperatures during operation to OFO and then returning to 1.3 GHz.

During this operation, the motor temperature was only warmed up by 11 K (see Fig. 4). In the figure it takes longer to reach OFO but this is due to the frequency measurements. This result shows that the copper heat sink attached to the stepper motor body can handle the heat load generated by the motor. The Fig. 4 shows how the stepper motor behaves during the OFO operation for CM1, the same behavior was observed in vCM.

The piezos experienced a large force when the OFO setpoint was reached. The piezo frequency sensitivities were measured before and after the OFO movement. No difference was observed, a more detailed study on the effects of multiple OFO cycles on the piezo is discussed in [4]. The piezo frequency sensitivity changes slightly when the stepper motor is at the OFO position. This is due to the larger force on the piezos.

The Higher order mode (HOM) couplers were out of specification at the OFO set point. Care must be taken when operating at this point since the HOMs will dissipate more power.

## CONCLUSION

Two LCLS-II-HE cryomodules were tested successfully at FNAL for OFO operation. It was demonstrated that both cryomodules were able to bring all 16 cavities to the OFO range. This was achieved by changing the tuner geometry and by making sure the cavity frequency was not out of the range of the tuner. During this large frequency detuning of the cavities the stepper motor was operated for roughly 2 hours. During this operation, the motors only warmed up by 11 K from the nominal value when it was static. No limit switches were tripped, and the 62 % specification to bring cavities to OFO was met.